# Generic Framework of Knowledge-Based Learning: Designing and Deploying of Web Application

Awais Khan Jumani[1(✉)], Anware Ali Sanjrani[2],
Fida Hussain Khoso[3], Mashooque Ahmed Memon[4],
Mumtaz Hussain Mahar[5], and Vishal Kumar[1]

[1] Faculty of Science and Technology, ILMA University, Karachi, Sindh, Pakistan
awaisjumani@yahoo.com, hit.vishal@outlook.com
[2] Department of Computer Science, University of Baluchistan, Quetta, Pakistan
anwar.csd@gmail.com
[3] Department of Basic Sciences, Dawood University of Engineering
and Technology, Karachi, Pakistan
fidahussain.khoso@duet.edu.pk
[4] Department of Computer Science, Benazir Bhutto Shaheed University,
Layari, Karachi, Pakistan
pashamorai786@gmail.com
[5] Department of Computer Science, Shah Abdul Latif University Khairpur,
Khairpur Mir's, Sindh, Pakistan
mhmahar@salu.edu.pk

**Abstract.** Learning technology was used as standalone software to install in a particular system, which needs to buy learning software of a particular subject. It was costly and difficult to search CD/DVD of the particular program in the market. Nowadays the trend of learning is changed and people are learning via the internet and it is known as Electronic Learning (E-learning). Several e-learning web applications are available which are providing more stuff about students and it fulfills requirements. The aim of this paper is to present a well-structured, user-friendly framework with the web application for e-learning, which does not need any subscription. The experiment was conducted with 691 students and teachers, the result shows 91.98% of participants were satisfied with the proposed E-learning system.

## 1 Introduction

The Internet was implemented and more established for educational organizations in the 1970s for communication system [1], instructors have been aware of its enormous perspective as a learning tool. Nowadays underdeveloped nations are excited to avail possibilities of online learning to convey actual cost, easily accessible and modern education of all ages and people backgrounds and geography. In the modern era, where the workers with good education are preferred as compared with workers with less knowledge and lifetime learning is realized as key to the nonstop achievement of modern civilization [2, 3]. E-learning is compared by many as the feasible solution to the problem providing the funds mandatory to smooth lifetime learning [4, 5].

However, present practices and theories of e-learning are not lucid, significance and scalable [6, 7]. Although most distinguish learning frameworks are available and it's likely to enhance prominent learning and experience of learning at all stages, people have realized its hitches and presently still too boundless to commit it seriously [8]. While several works have been revealed as well as written for e-learning strategy, but it still looks incapable to really express how, when or where eLearning should best be used [9, 10]. This paper shows that when it can be using E-Learning and what kind of eLearning should be used in our schools and colleges. Some successful results have been taken from different schools while using eLearning management. E-Learning considered as a globally that impacts much more of board learning [11, 12].

E-Learning activities collaborating technologies and communication system is more progress in the learning involvement [13]. It has been impending to renovate the way to learn and teach transversely the board [14]. It can increase values and enlarge participation in lifetime learning. It cannot swap lecturers and teachers, besides prevailing methods it can improve the quality and reach of their teaching and decrease the time consumed on administration. It can facilitate every beginner to accomplish his or her perspective and help to construct an educational workforce endowed to change. It makes imaginable a truly aspiring system for a future learning civilization.

## 1.1 Web Applications

Evolution of web application is using everywhere and these are dynamic websites, which is combination of server-side and client-side programming [15, 16]. Some of the facilities, for example, interacting with users, linking to back-end databases, and creating results to browsers. Web Application Frameworks are a group of many libraries programs, modules, and tools structured in an architecture system so it allows developers to construct and maintain complex web application program projects with their efficient methodology [17]. Some of the web application is updating programming and stimulate code reuses for common functions and classes. Those can be created dynamically website that can be taking JavaScript for making dynamically webpages. ASP.NET user can use object-oriented programming with C sharp, it can give a lot of facilities to the user for making practical or useful web application [18].

Entity Framework Fundamental is the newest version of the lightweight Entity Framework designed to work with .NET Fundamental applications. Suppose, ASP. NET Essential has been revised from the ground up and contains several new ways to perform. One of those is the introduction of the Track Graph technique for controlling multifarious data in disconnected situations such as Model View Controller (MVC) or Web API applications. The Track Graph technique is new in Entity Framework Essential and offers a simple way to repeat over a graph of objects that the perspective to begin tracking and to apply personalized code based on the type of entity and other benchmarks [19]. Early traditional web applications have not given the guarantee of and also additional plug-in required for viewing the lectures on flash player. Figure 1 shows a traditional application interface where videos or audio lectures are not available.

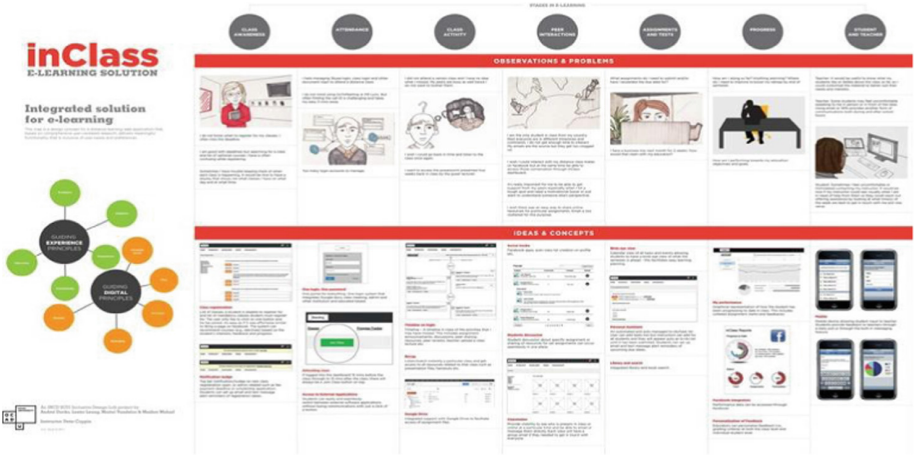

**Fig. 1.** Traditional web applications

## 2 Literature Review

Many researchers have provided research work regarding this work. Nagarajan and Jiji have introduced that online education has three major activities: Design, application and proper post-implementation assessments [19]. It reduces the cost and time efficiency, which is the main factor of our education and this framework have improved the quality of e-learning. But it gives the particular subjects knowledge environment and that efficiency.

Kumar et al. have proposed an E-learning framework for disabled persons whose arms are disabled to access systems. The technique of this framework is that disable student can use a computer by their voice control and learn by the e-learning software. ICT gives a lot of features for them but still, the framework has limitation for disabled persons to access [20].

Gopalan et al. have discussed in his research about the student of rural areas, which did not attend schools intermittently due to less source of income and working. E-learning education makes it possible if a student could not attend classes than they can learn these all thing from smartphones having e-learning app for students and they can interact with the teacher by using an application as like real environment [21]. Another research work given by Zhang and Goel, they have compared that e-learning technology usually used in higher education institutes. During the comparisons, they make two possibilities of results, first e-learning synthesized by identifying a relevant framework and second theoretical results. After getting results of e-learning it shows that it is more appropriate as compared to theoretical learning [22].

Similarly, Wu et al. were reviewed that e-learning is growing from 2005 and many of the research papers on their efficiency on their better effectiveness, a lot of the ratio of using e-learning in Croatia. It shows the relationship between user and perceptions of user characteristics; it gives limited atmosphere in web application and security issues concerned [23]. Similarly, Mosharraf et al. have applied the learning management system for the elementary schools and achieved successful results during the implementation. In the learning management system for kids, it gives more accurate results and students get more serious interest to learn through-out learning management system [24].

Further, Patel et al. has been explored that influence of e-learning in the life of learner that is a brilliant way to teach students using e-learning tools. Using e-learning tools they can enormously increase the learning process. The important remunerations of e-learning are that they can make it always on and update through connect the organizations [25]. Hence, Noesgaard and Ørngreen [26] have revealed that e-learning is more effective as compared to face to face learning methods. After the analysis of the student's perspective e-learning gives more good result and teachers should apply the method e-learning. Furthermore, Jumani et al. [27] have designed an application for kids to measure their mental approach using web application thus that application shows the interest of kids in learning with traditional and game-based learning using web application. He has taken their results which are overcome to traditional learning methodology.

## 3 Design and Development of Proposed Framework

Initially, the intranet environment is created for application testing. Inbounds of the firewall involved with the port exception, which enable to send and receive the request and response from the centralized server. Moreover, SQL Server has installed for storing, retrieving and manipulating records accompanied with Data Definition Language (DDL) and Data Manipulation Language (DML). Furthermore, C# integrated development environment along with ASP.Net technology has used for calling the classes. Namespaces and different libraries are used to accomplish the task. Cascade style sheet is played a crucial role in designing the application interface. Proposed application based on a few tabs which are given below with its description.

1. Video: This tab provides all the lectures of their subjects in the form of Videos and these can run on any media player or built-in flash player and these are special peculiarities. These videos also able to download and store on a computer drive. Once it reaches on end different thumbnail suggested the other related videos and these are similar to available media sites but different in functioning.
2. Documents: In this tab, more than 8000 doc or docx files have already stored and it is also related to their selected subjects. Document viewer control is used for getting a preview of the document and have feature, which enables to share this one with their friends. Moreover, the document can be stored and send via Bluetooth device to their mobiles.

3. PDF Document: This tab provides to students approximately 9000 pdf documents on different topics based on their subjects. Students are able to download the document and take a snapshot as well. Furthermore, all functionalities of zooming and rotating along with a quick search are integrated with the portable document viewer.

4. PowerPoint Presentations: Several pptx or ppt has already stored in that tab for providing different lectures slides with pictorial representation and it also enables to students to convert these presentations into a compact disc. Different themes and styles accompanied by slides have installed predefined and these are constant nobody can change it but only are able to download it or directly print through-out the printer. Users can also view these slides directly by clicking on the required topic or author name. For this, Microsoft office built-in components and libraries have been used from visual studio 2013 and some online other packages.

Some snapshots of the web application are given in Figs. 2, 3 and 4. Allah Dino Jumani (ADJ) E-learning application provide better platform for teachers and students In Fig. 2 user can first, sign-up an account and after it sent an activation link to the user email address. Once the user clicks on provided on link account successfully done with that process and the user can log in with a credential such as a user name and password. Moreover, easy and connect process with ADJ E-learning System gives notifications as well with the passage of time. After the login approvals, the user can easily access many different services.

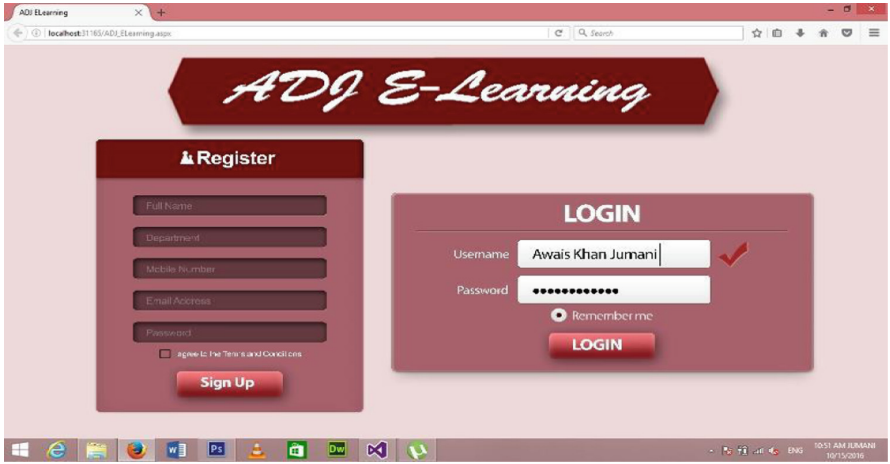

**Fig. 2.** Illustrates login and signup screen

In Fig. 3 user can see four tabs in front of their screen. In video tab user can easily watch any video of their concerned subject, it can be synchronized and bookmark of that video. Also, it can be downloaded directly and that can share it with the social network. This tab will offer some different aspect of e-learning any user can be upload video with concerned lectures and give online comments.

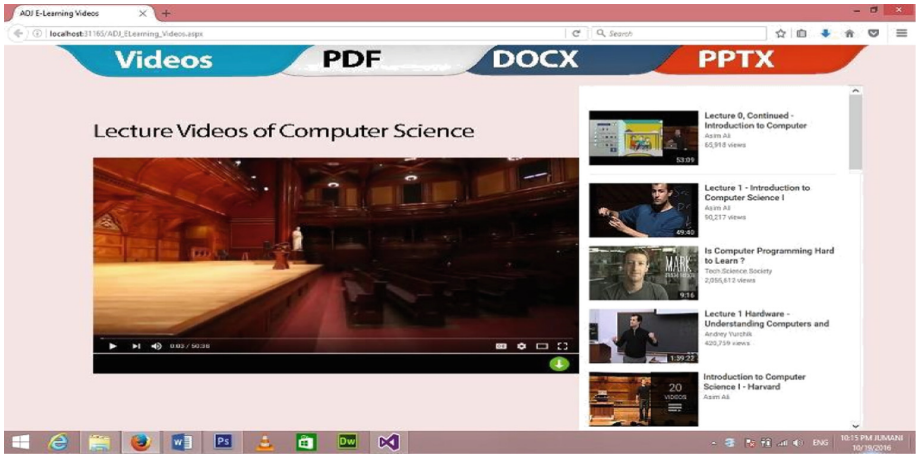

**Fig. 3.** Displays video lectures

In Fig. 4 user can see the pdf lectures of their related subject and it can be downloaded or directly view in Adobe Acrobat reader. It can be shared online in any social network with other users; every pdf lecture is updated with new technologies and close with authenticating references.

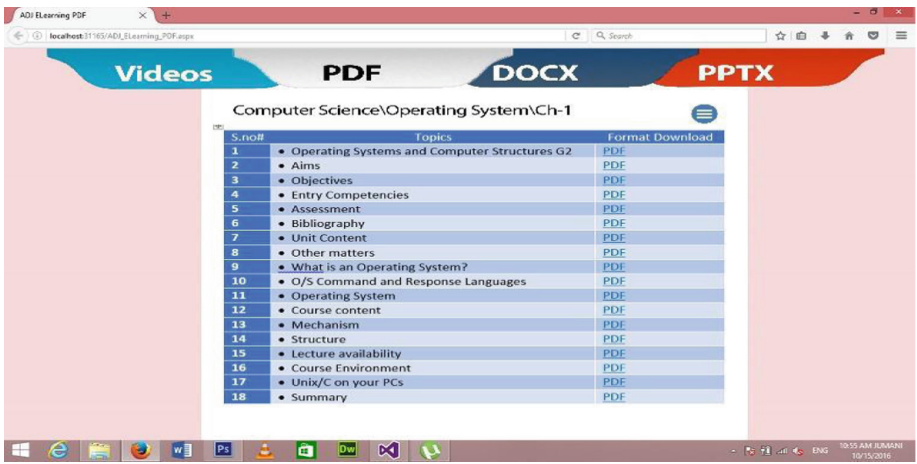

**Fig. 4.** Displays PDF lecture topics

In the proposed framework, initially, the user should have their email address of an account. The user can sign up their account, it requests to server and server sends the activation account link to their given email address. If an account activation link acknowledges from the given email account, it will be activated. After confirmation user can sign in from the inbox of ADJ E-Learning web application similar to Yahoo and Gmail and then user interact with all present functionalities. If a link which has sent will not confirm by the user it will expire and ADJ has resent option is available for next use. The flowchart shows the overall process of sign up to log out of a user in Fig. 5.

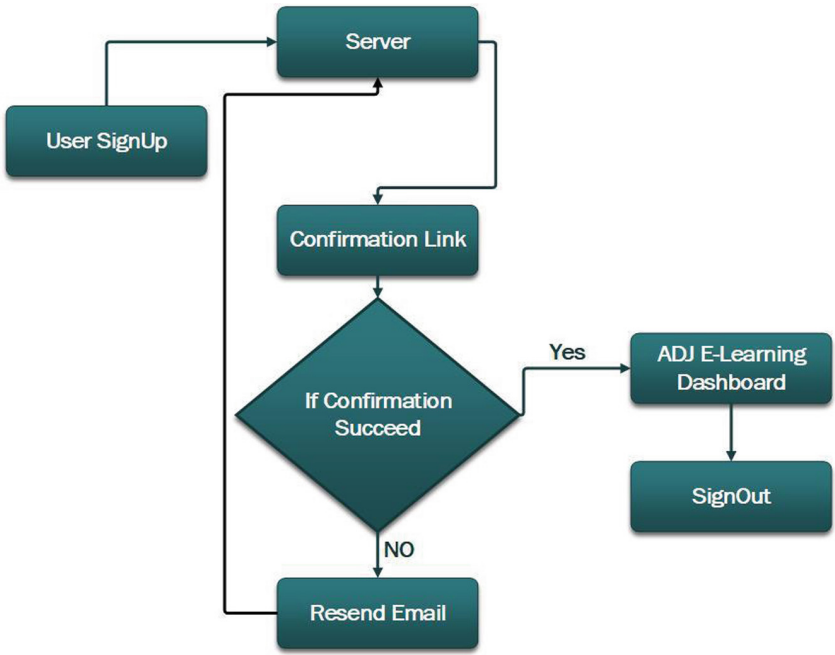

**Fig. 5.** Framework of ADJ E-learning

## 4  Results

For the evaluation of the proposed framework, we have selected 691 teachers and students simultaneously and achieved positive results from 4 major peculiarities (1) Convenient Environment, (2) Secure, (3) Useful and (4) Reliable. Overall, selected responders have given in Table 1.

**Table 1.** Selected responders from Shah Abdul Latif University

| Teacher | |
|---------|-----|
| Male | 67 |
| Female | 54 |
| Student | |
| Male | 235 |
| Female | 335 |

Table 1 shows that 67 male (M) teachers were chosen from SALU and given 60 responds from M teachers accompanied with 89.55% positive satisfaction response. From the side of female (F) teachers, 54 were chosen and 49 has given respond with positive satisfaction accompanied with 90.74%.

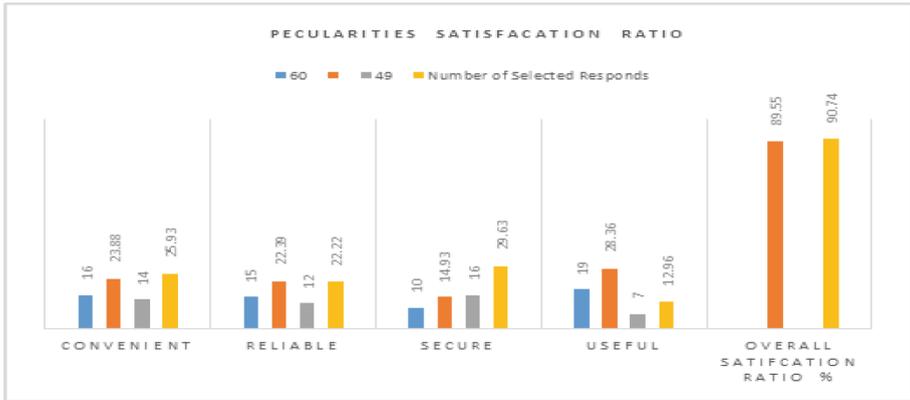

**Fig. 6.** Teachers peculiarities satisfaction ratio

On the other side, 235 M students have chosen and 225 students have given respond along with 95.74% satisfaction, 335 F students were chosen and 295 has to respond along with 88.06% satisfaction.

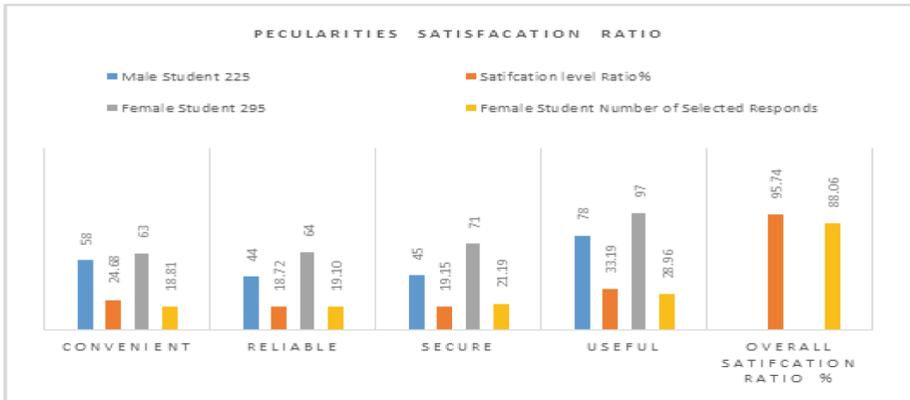

**Fig. 7.** Student peculiarities satisfaction ratio

Figures 6 and 7 represents that among the four aspects, respondents have given reviews that, presented application more convenient, useful for students and will give fruitful results to up-coming generation.

During testing of the proposed application by selected responded it is observed, the majority of responded were interested to download videos in contrast to a text document. Figure 8 illustrates focused contents while testing.

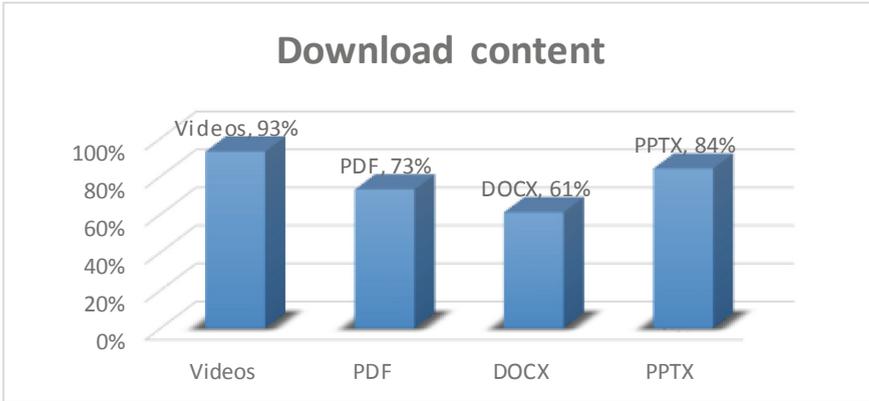

**Fig. 8.** Percentages of downloaded content

## 5    Conclusion

We have presented a convenient framework wrapped with several peculiarities for providing academic stuff regarding their subjects. There are several e-learning platform available, based on the users expectation features are varied. In our platform there is no need for any subscription by using this application student can download several lectures presentation without any cost. Technology grows day by day and also educational institutes use electronic learning system. In future this E-Learning application will be used in institutes, it will give a lot of facilities to students in their studies and with the help of e-learning, technology teacher can easily teach students with graphical representations. Some of the students can take advantages from E-learning educational system which did not give much time to their institutes and they can easily learn at home through the e-learning web application.